\documentclass[aps,
		prd,
		reprint,
		twocolumn,
		superscriptaddress,
		shortbibliography,
		nofootinbib,
		floatfix,
		notitlepage
		]{revtex4-1}
\pdfoutput=1

\usepackage{amsmath,amssymb}
\usepackage{graphicx,accents}
\usepackage{graphicx,epstopdf}
\usepackage[usenames,dvipsnames]{xcolor}
\usepackage{slashed}
\usepackage[normalem]{ulem}
\usepackage{bm}
\usepackage{bbm}
\usepackage{bbold}
\usepackage{float}

\usepackage[printwatermark]{xwatermark}

\usepackage{hyperref}
\hypersetup{colorlinks=true}

\usepackage{tabularx}

\newcommand{\be}{\begin{equation}}
\newcommand{\ee}{\end{equation}}
\newcommand{\ba}{\begin{array}}
\newcommand{\ea}{\end{array}}
\newcommand{\bea}{\begin{eqnarray}}
\newcommand{\eea}{\end{eqnarray}}
\newcommand{\bd}{\begin{displaymath}}
\newcommand{\ed}{\end{displaymath}}

\newcommand{\trm}[1]{\textrm{#1}}

\newcommand{\vphi}{\varphi}

\newcommand{\ud}{\mathrm{d}}
\newcommand{\LCm}{{\scriptscriptstyle -}} 
\newcommand{\LCp}{{\scriptscriptstyle +}}

\newcommand{\LCperp}{{\scriptscriptstyle \perp}}
\newcommand{\bi}{\begin{itemize}}
\newcommand{\ei}{\end{itemize}}

\makeatletter
\DeclareRobustCommand{\cev}[1]{%
  \mathpalette\do@cev{#1}%
}
\newcommand{\do@cev}[2]{%
  \fix@cev{#1}{+}%
  \reflectbox{$\m@th#1\vec{\reflectbox{$\fix@cev{#1}{-}\m@th#1#2\fix@cev{#1}{+}$}}$}%
  \fix@cev{#1}{-}%
}
\newcommand{\fix@cev}[2]{%
  \ifx#1\displaystyle
    \mkern#23mu
  \else
    \ifx#1\textstyle
      \mkern#23mu
    \else
      \ifx#1\scriptstyle
        \mkern#22mu
      \else
        \mkern#22mu
      \fi
    \fi
  \fi
}

\newcommand*\xbar[1]{%
  \hbox{%
    \vbox{%
      \hrule height 0.5pt 
      \kern0.2ex
      \hbox{%
        \kern-0.15em
        \ensuremath{#1}%
        \kern-0.15em
      }%
    }%
  }%
}

\newcommand{\vtheta}{\vartheta}
\graphicspath{ {./figures/} }

\makeatletter

\newcommand{\Rmnum}[1]{\expandafter\@slowromancap\romannumeral#1@}
\makeatother
\begin{document}
\title{Optimal photon polarization toward the observation of the nonlinear Breit-Wheeler pair production}
\author{Yunquan Gao}
\affiliation{College of Physics and Optoelectronic Engineering, Ocean University of China, Qingdao, Shandong, 266100, China}

\author{Suo Tang}
\email{tangsuo@ouc.edu.cn}
\affiliation{College of Physics and Optoelectronic Engineering, Ocean University of China, Qingdao, Shandong, 266100, China}


\begin{abstract}
We investigate the optimization of the photon polarization to increase the yield of the Breit-Wheeler pair production in arbitrarily polarized plane wave backgrounds.
We show that the optimized photon polarization can improve the positron yield by more than $20\%$ compared to the unpolarized case, in the intensity regime of current laser-particle experiments.
The seed photon's optimal polarization is resulting from the polarization coupling with the polarization of the laser pulse.
The compact expressions of the coupling coefficients in both the perturbative and nonperturbative regime are given.
Because of the evident difference in the coupling coefficients for the linear and circular polarization components,
the seed photon's optimal polarization state in an elliptically polarized laser background, deviates considerably from the orthogonal state of the laser polarization.
\end{abstract}
\maketitle
%
%
\section{Introduction}
The production of electron-positron pair in the collision of two high-energy photons, now referred to as the linear Breit-Wheeler process (LBW), was first proposed in 1930s~\cite{PhysRev.46.1087}.
The production yield depends not only on the photons' dynamical parameters, but also on the relative polarization of the two photons~\cite{PhysRev.46.1087,Baier2002hep,PRL2021_052302}.

With the improvement of the laser intensity, the decay of a single high-energy photon into a pair of electron and positron in the collision with an intense laser pulse,
which is often referred to as the nonlinear Breit-Wheeler (NBW) pair production~\cite{Reiss1962,RMPPiazza,2021arXiv210702161G,Fedotov2022ely},
has been measured in the multiphoton perturbative regime via the landmark E144 experiment more than two decades ago~\cite{PhysRevLett.79.1626,PhysRevD.60.092004} and been broadly studied within different type of laser fields~\cite{nikishov64,Heinzl:2010vg,PRA2012_052104,PRL2012_240406,PRA2013_062110,PhysRevD.93.045010,PhysRevA.88.052125,JANSEN201771,PhysRevD.98.036022,Anton19PRD125018,PhysRevD.101.016006,PhysRevA.101.042508,TangPRA2021,TangPRD096019}.
The dependence of the NBW process on the polarisation state of the seed photon has also been partially investigated in the current literature~\cite{ivanov2005complete,katkov2012production,PRL2020_014801,chen2022electron,PRD076017,Titov2020,SeiptPRA052805,Tang2022PRD}, in which the laser backgrounds are commonly specified with the pure linear and/or circular polarization,
and 
the production yield could be considerably improved/suppressed when the polarization of the seed photon is set to be orthogonal/parallel to that of the background field~\cite{Titov2020,SeiptPRA052805,Tang2022PRD}.
However, in an arbitrarily polarized laser background, how to assign the photon polarization to acquire the maximal production yield has not been clearly investigated.

In the LBW process, the polarization dependence of the production is resulting from the polarization coupling between the two high-energy photon~\cite{PhysRev.46.1087,Baier2002hep,PRL2021_052302}.
However, how the polarization of the seed photon couples with that of the laser pulse (or multiple laser photons) in the NBW process is still not clear.
In this paper, we concentrate on the properties of the polarization coupling between the seed photon and the laser pulse and reveal the optimal polarization of the seed photon for the maximal yield of the NBW process in arbitrarily polarized laser backgrounds.
We find that the linear and circular polarization component of the seed photon couple with the corresponding component of the laser polarization with the quite different coefficients, and thus in an elliptically polarized laser pulse, the optimal polarization state of the seed photon deviates considerably from the orthogonal state of the laser polarization.

The study of the optimal photon polarization for the maximal production yield is partly motivated by the upcoming high-energy laser-particle experiment, \emph{i.e.}, \mbox{LUXE} at \mbox{DESY}~\cite{Abramowicz:2021zja,Borysova2021,macleod2021theory,Jacobs:2021fbg} and \mbox{E320} at \mbox{SLAC}~\cite{slacref1,naranjo2021pair,E320_2021,Meuren:2020nbw}
in which beams of photons with the energy $O(10~\trm{GeV})$ are generated to collide with laser pulses with the intermediate intensity $\xi\sim O(1)$, and one of their main goals is to detect the NBW process in the transition regime from the perturbative to the non-perturbative regime~\cite{macleod2021theory,Jacobs:2021fbg}, where $\xi$ is the classical nonlinearity parameter for laser intensity.
In this planned intensity regime, the production yield could be enhanced/suppressed considerably by the photon polarization effect~\cite{PRD076017,Tang2022PRD}.

The paper is organised as follows. The theoretical model and relevant parameters are introduced in \mbox{Sec.~\ref{Sec_Theo}}.
In Sec.~\ref{Sec_LBW}, we first explore the perturbative intensity regime and discuss the photon polarization coupling in the LBW process,
and then, we go to the non-perturbative intensity regime to discuss the polarization coupling between the seed photon and the laser pulse in the NBW precess in Sec.~\ref{Sec_NBW}. At the end, we conclude in Sec.~\ref{sec_conc}.
In following discussions, the natural units $\hbar=c=1$ is used, and the fine structure constant is $\alpha=e^2\approx1/137$.

\section{Theoretical model}~\label{Sec_Theo}
We consider the typical scenario in the modern-day laser-particle experiments in which a beam of high-energy photons interacts with an intense laser pulse in the geometry close to the head-on collision.
The laser pulse is modelled as a plane wave with scaled vector potential $a^{\mu}(\phi)=|e| A^{\mu}(\phi)$ depending only on the laser phase $\phi=k\cdot x$, where $k^{\mu}=\omega(1,0,0,-1)$ is the laser wave vector, $\omega$ is the central frequency of the laser pulse and $|e|$ is the charge of the positron.
This plane wave background is a good approximation for collisions between high-energy particles and weakly focussed pulses~\cite{nikishov64,DiPiazza2015PRA,DiPiazza2016PRL,DiPiazza2017PRA032121,DiPiazza2021PRD076011}.
The collision is characterized by the energy parameter $\eta = k\cdot \ell/m^{2}$ and laser intensity parameter $\xi$, where $\ell^{\mu}$ is the photon momentum and $m$ is electron rest mass.

The total yield of the NBW pair production from a polarized seed photon is given as~\cite{Tang2022PRD}:
\begin{align}
{P}=&\frac{\alpha}{(2\pi\eta)^2}\int \frac{\ud s}{ts}\int\ud^{2} \bm{r}\iint \ud \phi_{1}\ud \phi_{2}~ e^{i\int_{\phi_{2}}^{\phi_{1}}\ud\phi'\frac{\ell\cdot\pi_{q}(\phi')}{m^{2}\eta t}}\nonumber\\
&\left\{h_s \bm{\Delta}^{2}/2 + 1 -ih_s\Gamma_{3} \bm{w}(\phi_{1}) \times \bm{w}(\phi_{2})\right.\nonumber\\
&      -\Gamma_{1} \left[w_{x}(\phi_{1}) w_{x}(\phi_{2}) - w_{y}(\phi_{1}) w_{y}(\phi_{2})\right]\nonumber\\
&\left.-\Gamma_{2} \left[w_{x}(\phi_{1}) w_{y}(\phi_{2}) + w_{y}(\phi_{1}) w_{x}(\phi_{2})\right]\right\}\,,
\label{NBW probability}
\end{align}
where $\bm{\Delta}=i\left[\bm{a}^{\LCperp}(\phi_1)-\bm{a}^{\LCperp}(\phi_2)\right]/m$, \mbox{$h_s=(s^2+t^2)/(2st)$}, $\bm{w}(\phi)= \bm{r}-\bm{a}^{\LCperp}(\phi)/m$, and $\bm{w}(\phi_{1}) \times \bm{w}(\phi_{2}) = w_{x}(\phi_{1}) w_{y}(\phi_{2}) - w_{y}(\phi_{1})w_{x}(\phi_{2})$,
$s=k\cdot q/k\cdot \ell$ ($t=1-s$) is the fraction of the light front momentum taken by the produced position (electron),
and $\bm{r}=(\bm{q}^{\LCperp}-s\bm{\ell}^{\LCperp})/m$ denotes the transverse momenta of the positron,
and $\pi_q(\phi)$ is the positron's instantaneous momentum in the laser pulse:
\[\pi^{\mu}_q(\phi)=q^{\mu}-a^{\mu}(\phi)+\frac{q\cdot a(\phi)}{k\cdot q}k^{\mu}-\frac{a^2(\phi)}{2k\cdot q}k^{\mu}\,.\]

The polarization of the seed photon is comprehensively described with the classical Stokes parameters $(\Gamma_1,~\Gamma_2,~\Gamma_3)$~\cite{landau4,Wiley3}:
$\Gamma_1$ ($\Gamma_2$) is the degree of linear polarization indicating the preponderance of the polarization in the $\varepsilon_{x}$ state ($\varepsilon_{45^{\circ}}$ state) over that in the $\varepsilon_{y}$ state ($\varepsilon_{135^{\circ}}$ state), and
$\Gamma_3$ is the degree of circular polarization giving the preponderance of the polarization in the $\varepsilon_{\LCp}$ state over that in the $\varepsilon_{\LCm}$ state.
The polarization basis is given as
\begin{align}
\varepsilon^{\mu}_{x}&=\epsilon^{\mu}_{x}-\frac{\ell\cdot \epsilon_{x}}{k\cdot \ell}k^{\mu}\,,~~~
\varepsilon^{\mu}_{y}=\epsilon^{\mu}_{y}-\frac{\ell\cdot \epsilon_{y}}{k\cdot \ell}k^{\mu}\,,\nonumber\\
\varepsilon^{\mu}_{\psi}&=\epsilon^{\mu}_{\psi}-\frac{\ell\cdot \epsilon_{\psi}}{k\cdot \ell}k^{\mu}\,,~~~
\varepsilon^{\mu}_{\pm}=\epsilon^{\mu}_{\pm}-\frac{\ell\cdot \epsilon_{\pm}}{k\cdot \ell}k^{\mu}\,,\nonumber
\end{align}
where $\epsilon^{\mu}_{x}=(0,1,0,0)$, $\epsilon^{\mu}_{y}=(0,0,1,0)$ and $\epsilon_{\psi} = \epsilon_{x}\cos\psi + \epsilon_{y}\sin\psi$, $\epsilon_{\pm} = (\epsilon_{x} \pm i\epsilon_{y})/\sqrt{2}$.
For fully polarized photon beams, the Stokes parameters satisfy $\Gamma_1^2+\Gamma_2^2+\Gamma_3^2=1$
and for partially polarized photon beams, $\Gamma_1^2+\Gamma_2^2+\Gamma_3^2<1$.
The full definition of the photon Stokes parameters $(\Gamma_1,~\Gamma_2,~\Gamma_3)$ can be found in Ref.~\cite{Tang2022PRD}.

Based on~(\ref{NBW probability}), the total yield of the NBW process can be phenomenologically given as
\begin{align}
\trm{P}=n_{0} + \Gamma_{1} n_{1} + \Gamma_2 n_{2} + \Gamma_{3}n_{3}\,,
\label{Eq_T_Yiled}
\end{align}
where $n_{0}$ is the unpolarized contribution independent on the photon polarization \hbox{$(\Gamma_{1,2,3}=0)$}~\cite{TangPRA2021,TangPRD096019},
and $n_{1,2,3}$ denote the contributions coupling to the polarization of the seed photon.
As one can simply infer, to maximize the production yield,
\begin{align}
\trm{P}_{m}=n_0+n_{p}\,,
\label{Eq_M_Yiled}
\end{align}
the photon polarization should be selected as
\begin{align}
(\Gamma_{1}, \Gamma_{2}, \Gamma_{3})= (n_{1}, n_{2}, n_{3})/(n_{1}^2 + n_{2}^2 + n_{3}^2)^{1/2}\,,
\label{Eq_opti_polar}
\end{align}
which prompts the existence of the optimal photon polarization for the specified laser pulse and collision parameter $\eta$ to achieve the maximal production yield,
where $n_{p}=(n_1^2+n_2^2+n_3^2)^{1/2}$ is the maximal contribution from the photon polarization.
However, if reverse the optimal polarization of the seed photon, \emph{i.e.} $\Gamma_{1,2,3}\to -\Gamma_{1,2,3}$, the pair production would be largely suppressed.

\section{linear Breit-Wheeler process}~\label{Sec_LBW}
One may realize that the polarization contribution $\Gamma_{i}n_{i}$ in~(\ref{Eq_T_Yiled}) comes from the polarization coupling between the seed and laser photons, and thus the optimal photon polarization~(\ref{Eq_opti_polar}) depends on the polarization of the laser photons.
To manifest this polarization coupling effect, we resort to the perturbative approximation of~(\ref{NBW probability}), which is often referred to as the LBW process, by expanding the integrand in~(\ref{NBW probability}), keeping only $\mathcal{O}(\xi^2)$ terms and integrating over $s$,
\begin{align}
{P}_{\ell}=&\frac{\pi\alpha^2\lambdabar^{2}_{e}}{2}\int_{\nu_{*}}^{+\infty}\ud{\nu}~D(\nu) \nonumber\\
 &\left\{ \Xi +\kappa_{c}\Gamma_{3} \varsigma_{3}(\nu) + \kappa_{l}[\Gamma_{1} \varsigma_{1}(\nu) + \Gamma_{2} \varsigma_{2}(\nu)]\right\}\,,
 \label{Perturbative probability}
\end{align}
where $\nu_{*}=2/\eta$ is the frequency threshold of the laser photon required to trigger the pair production, \mbox{$D(\nu) = \nu |\tilde{\bm{a}}(\nu)|^{2} /(4\pi^{2}\alpha\lambdabar^{2}_{e} m^{2})$} is the (areal) number density of the laser photon with the frequency $\nu\omega$, \mbox{$\lambdabar_{e}=1/m=386.16~\trm{fm}$} is the electron's reduced Compton wavelength, \mbox{$\tilde{\bm{a}}(\nu)=\int \ud \phi [a_{x}(\phi),~a_{y}(\phi)]\exp(iv\phi)$}, and
\begin{align}
  \begin{aligned}
\varsigma_{1}(\nu)&=\frac{|\tilde{a}_{x}(\nu)|^2-|\tilde{a}_{y}(\nu)|^2}{|\tilde{\bm{a}}(\nu)|^2},&\\
\varsigma_{2}(\nu)&=\frac{\tilde{a}^{*}_{x}(\nu)\tilde{a}_{y}(\nu)+\tilde{a}_{x}(\nu)\tilde{a}_{y}^{*}(\nu)}{|\tilde{\bm{a}}(\nu)|^2},&\\
\varsigma_{3}(\nu)&=i\frac{\tilde{a}_{x}(\nu)\tilde{a}_{y}^{*}(\nu) - \tilde{a}^{*}_{x}(\nu)\tilde{a}_{y}(\nu)}{|\tilde{\bm{a}}(\nu)|^2},
\end{aligned}
\label{Eq_photon_Stokes}
\end{align}
are the classical Stokes parameters of the laser photon $\nu\omega$~\cite{Wiley3}, satisfying \mbox{$\varsigma^{2}_{1}(\nu)  + \varsigma^2_{2}(\nu) + \varsigma^2_{3}(\nu)=1$}. Similar as the seed photon, $\varsigma_{1,2,3}(\nu)$ characterize the polarization property of the laser photon:
$\varsigma_{1}(\nu)$ [$\varsigma_{2}(\nu)$] describes the preponderance of the $\epsilon_{x}$ ($\epsilon_{45^{\circ}}$)-linear polarization over the $\epsilon_{y}$ ($\epsilon_{135^{\circ}}$)-linear polarization,
and $\varsigma_{3}(\nu)$ denotes the preponderance of the $\epsilon_{+}$-circular polarization over the $\epsilon_{-}$-circular polarization. The parameter
\begin{align}
\Xi=(1-\beta^2)\left[(3-\beta^4)\ln\left(\frac{1+\beta}{1-\beta}\right) - 2\beta(2-\beta^2)\right]
\end{align} is the contribution from unpolarized photons~\cite{PRD_16_286,Greiner}, and
\begin{align}
\kappa_{c}&=2(1-\beta^2)\left[\ln\left(\frac{1+\beta}{1-\beta}\right)-3\beta\right]\,,\\
\kappa_{l}&=-\frac{(1-\beta^2)^3}{2}\left[\ln\left(\frac{1+\beta}{1-\beta}\right) + \frac{2\beta}{1-\beta^2}\right]
\end{align}
are, respectively, the circular- and linear-polarization coupling coefficients, and indicate the amplitude of the contributions from each kind of polarization coupling between the seed and laser photons,
where $\beta=(1- \nu_{*}/\nu)^{1/2}$ is actually the normalized velocity of the produced particles in the center-of-mass frame.

\begin{figure}[t!!!]
\includegraphics[width=0.45\textwidth]{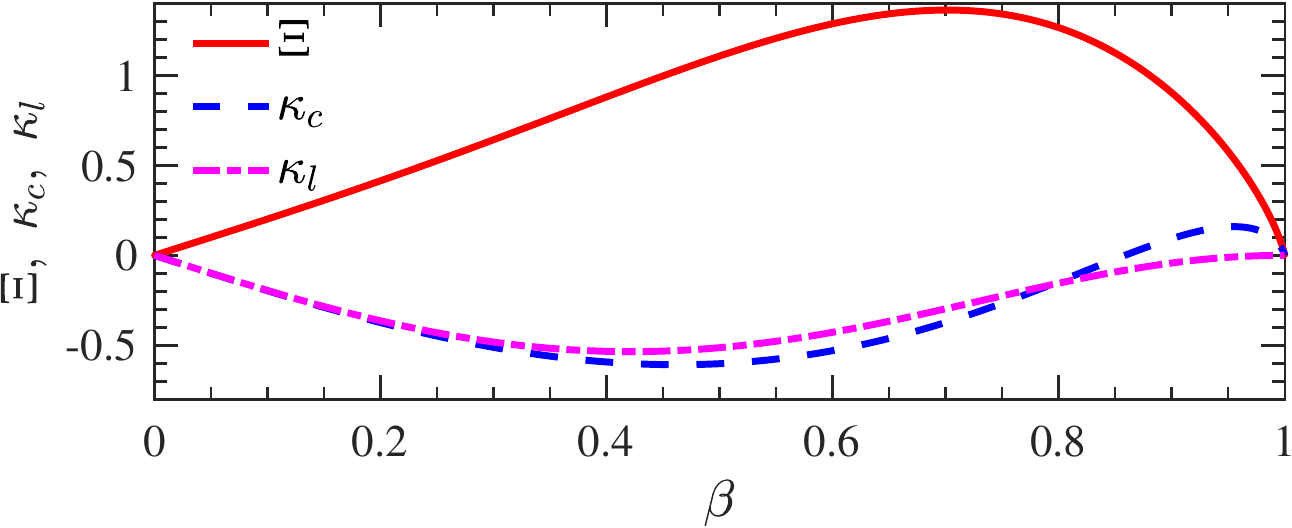}
   \caption{Comparison between the polarization contributions $\kappa_{c,l}$ and the unpolarized contribution $\Lambda$ with the change of the parameter $\beta$ in the linear Breit-Wheeler process. $\beta$ is defined in the text and can be simply understood as the normalized velocity of the produced particles in the center-of-mass frame.}
   \label{Fig1_pert}
\end{figure}

In~(\ref{Perturbative probability}),~we can clearly see the contributions from the polarization coupling between the seed and laser photons. To maximize the polarization contribution in the LBW process, the polarization of the seed photon is optimized, based on the polarization of the laser photon, as
\begin{align}
(\Gamma_{1},~\Gamma_{2},~\Gamma_{3})=\hat{\kappa}_{l}\frac{[\varsigma_{1}(\nu),~\varsigma_{2}(\nu),~\sigma \varsigma_{3}(\nu)]}{[\varsigma_{1}^2(\nu) + \varsigma_{2}^2(\nu) + \sigma^2  \varsigma_{3}^2(\nu)]^{1/2}}\,,
\label{Eq_opti_polar_pert}
\end{align}
where $\sigma_{l} = \kappa_{c}/\kappa_{l}$, and $\hat{\kappa}_{l}$ is the sign of $\kappa_{l}$.
As we can also see in~(\ref{Perturbative probability}), the two sets of linear polarization have the identical coupling coefficient $\kappa_{\ell}$, because of the symmetry by rotating the linear polarization axis for $45^{\circ}$.
This identity results in the orthogonality between the linear polarization components of the seed and laser photons as \mbox{$(\Gamma_{1},~\Gamma_{2}) \sim -[\varsigma_{1}(\nu),~\varsigma_{2}(\nu)]$} as shown in~(\ref{Eq_opti_polar_pert}) where \mbox{$\hat{\kappa}_{l}=-1$} is obtained in Fig.~\ref{Fig1_pert}.

In Fig.~\ref{Fig1_pert}, the unpolarized contribution $\Xi$ and the polarization coupling coefficients $\kappa_{l,c}$ are presented with the change of the parameter $\beta$.
As shown, the polarization contributions are indeed appreciable compared with the unpolarized contribution, especially in the low-energy region $\beta < 0.2$, where \mbox{$\Xi\approx-\kappa_{c} \approx -\kappa_{l}$} and the energy of the laser photon is close to the frequency threshold $\nu\to \nu_{*}$.
With the proper photon polarization, the production could be doubled if $\bm{\Gamma}\cdot\bm{\varsigma}(\nu)\to-1$ or completely suppressed if $\bm{\Gamma}\cdot\bm{\varsigma}(\nu)\to 1$.
Similar as the variation of the unpolarized contribution $\Xi$ with \mbox{$\beta\in(0,1)$}~\cite{Greiner},
the amplitude of the coupling coefficients $\kappa_{c,l}$ increase from zero at $\beta=0$ to the maximum at around $\beta\approx 0.45$ and then fall off again to zero at $\beta=1$.
In the region of $\beta<0.4$, the two kind of polarization have the same coupling coefficient, $\kappa_{c}\approx\kappa_{l}$.
This means that, to acquire the maximal polarization contribution, the seed photon should be fully polarized in the state orthogonal to that of the laser photon, \emph{i.e.} $(\Gamma_{1},~\Gamma_{2},~\Gamma_{3})=-[\varsigma_{1}(\nu),~\varsigma_{2}(\nu),~\sigma \varsigma_{3}(\nu)]$ with $\sigma_{l}\approx 1$ in~(\ref{Eq_opti_polar_pert}).
However, in the higher-energy region with $\beta>0.4$, the difference between $\kappa_{c}$ and $\kappa_{l}$ becomes considerable, which implies that the highest production yield is acquired from the seed photon polarized in the state deviating from the orthogonal state of the laser photon. Especially in the extremely high-energy region with $\beta>0.95$ in which $\kappa_{l}$ is close to zero and $\kappa_{c}$ becomes positive and dominates the polarization contribution, the highest yield appears when the seed and laser photons have the pure circular polarization parallel to each other.

We now know that the polarization coupling between the two photons in the LBW process could contribute considerably to the production yield and the polarization contributions $n_{1,2,3}$ in~(\ref{Eq_T_Yiled}) are proportional to the Stoke parameters of the laser photon as $n_{1,2}\sim D(\nu)\kappa_{l}\varsigma_{1,2}(\nu)$ and $n_{3}\sim D(\nu)\kappa_{c}\varsigma_{3}(\nu)$ with the coupling coefficients $\kappa_{l,c}$ depending only on the dynamic parameter $\beta$ in the perturbative regime $\xi\ll 1$.
While in the upcoming laser-particle experiments~\cite{Abramowicz:2021zja,naranjo2021pair},
the laser intensity has increased to the magnitude of $\xi\sim \mathcal{O}(1)$, in which the Breit-Wheeler pair production is in the transition regime from the perturbative to the non-perturbative regime,
a high number of laser photons would be involved to satisfy the energy threshold in the center-of-mass frame,
and the NBW process would dominate the pair production.
The polarization contributions would, therefore, come from the polarization coupling with the laser pulse, \emph{i.e.} multiple laser photons, but not with a single laser photon, and the coupling coefficients would depend also on the laser intensity and field ellipticity.

\section{nonlinear Breit-Wheeler process}~\label{Sec_NBW}
In this section, we consider the NBW process stimulated by a high-energy photon in the collision with the laser pulse in the intermediate intensity region $\xi\sim \mathcal{O}(1)$. This is the typical setup for the upcoming laser-particle experiment in LUXE~\cite{Abramowicz:2021zja,Jacobs:2021fbg}.
To show the polarization effect clearly, we fix the energy parameter $\eta$ and adjust the relative polarization of the seed photon and laser pulse.

The background laser field is expressed as
\begin{align}
a^{\mu}(\theta,\phi)=m\xi~\trm{Re}\left\{\left[0,a_x(\theta),a_y(\theta),0\right]e^{-i\phi}\right\}f(\phi),
\label{Vector potential}
\end{align}
where $\trm{Re}\left\{\cdot\right\}$ means the real part of the argument, $a_x(\theta)=\cos \theta - i \delta \sin \theta$, $a_y(\theta)=\sin \theta + i\delta \cos \theta $.
\mbox{$\delta\in [-1, 1]$} characterizes not only the rotation of laser field: $\delta/|\delta|=1$ means the left-hand rotation and $\delta/|\delta|=-1$ is right-hand rotation, but also the ellipticity $|\delta|$ of the laser pulse: $|\delta|=0,~1$ corresponds, respectively, to the linearly and circularly polarized laser background and \mbox{$0<|\delta|<1$} gives a laser pulse with the elliptical polarization.
The semi-major axis of the elliptical laser field is along ($\cos\theta,~\sin\theta$) with the deflection angle $\theta\in[-\pi,~\pi]$ in the transverse plane. 
$f(\phi)$ depicts the envelope of the laser pulse.
The polarization of the laser field could also be described with the classical Stokes parameters $(\varsigma_1,~\varsigma_2,~\varsigma_3)$~\cite{Wiley3} as
\begin{align}
\begin{aligned}
\varsigma_{1}&=\frac{|a_{x}|^{2} - |a_{y}|^{2}}{|a_{x}|^{2} + |a_{y}|^{2}}=\frac{1-\delta^{2}}{1+\delta^2} \cos{2\theta}\,,&\\
\varsigma_{2}&=\frac{a^{*}_{x}a_{y} + a_{x}a^{*}_{y}}{|a_{x}|^{2} + |a_{y}|^{2}}=\frac{1-\delta^{2}}{1+\delta^2} \sin{2\theta}\,,&\\
\varsigma_{3}&=i\frac{a_{x}a^{*}_{y}-a^{*}_{x}a_{y}}{|a_{x}|^{2} + |a_{y}|^{2}}=\frac{2\delta}{1+\delta^2}\,.
\end{aligned}
\label{Eq_laser_Stokes}
\end{align}
where $\varsigma^{2}_{1} + \varsigma^{2}_{2}+ \varsigma^{2}_{3}=1$.
The total linear polarization degree of the laser pulse is given as \mbox{$\varsigma_{l} = (\varsigma^{2}_{1} + \varsigma^{2}_{2})^{1/2} =(1-\delta^{2})/(1+\delta^2)$}, and the laser's circular polarization degree is given by $\varsigma_{3}$.
The equivalence between the laser Stokes parameters~(\ref{Eq_laser_Stokes}) and those of the laser photon~(\ref{Eq_photon_Stokes}) can be seen when we consider a relatively long laser pulse with the slowly varying envelope \mbox{$f'(\phi)\approx 0$} and $|\tilde{f}(\nu+1)|\ll |\tilde{f}(\nu-1)|$ at \mbox{$\nu\geq 1$}~\cite{TangPRD096019}.
The frequency components of the laser pulse can be written approximately as~\mbox{$\tilde{a}^{\mu}(\nu)\approx m\xi/2~\left[0,a_x(\theta),a_y(\theta),0\right] \tilde{f}(\nu-1)$} and therefore $\varsigma_{i}\approx \varsigma_{i}(\nu)$ with $i=1,2,3$.

\subsection{Numerical results}
To show the importance of polarization contributions and their dependence on the corresponding laser Stokes parameters,
we first present the numerical results for the NBW process stimulated by a $16.5~\trm{GeV}$ photon in the head-on collision with the laser pulse in the intermediate intensity region $\xi\sim \mathcal{O}(1)$. 
The pulse envelope is given as $f(\phi)=\cos^2[\phi/(4\sigma)]$ in $\left|\phi\right|<2\pi \sigma$ and $f(\phi)=0$ otherwise, where $\sigma=8$.
The calculations have been done with the laser central frequency $\omega=4.65~\trm{eV}$, as an example, which is the third harmonic of the normal laser with the wavelength $\lambda=0.8~\mu\trm{m}$.
For the detail calculation of~(\ref{NBW probability}), one can refer to the presentation in Ref.~\cite{TangPRA2021} and the analogous calculation in Ref.~\cite{BenPRA2020} for the polarized nonlinear Compton scattering.

\begin{figure}[t!!!]
\includegraphics[width=0.45\textwidth]{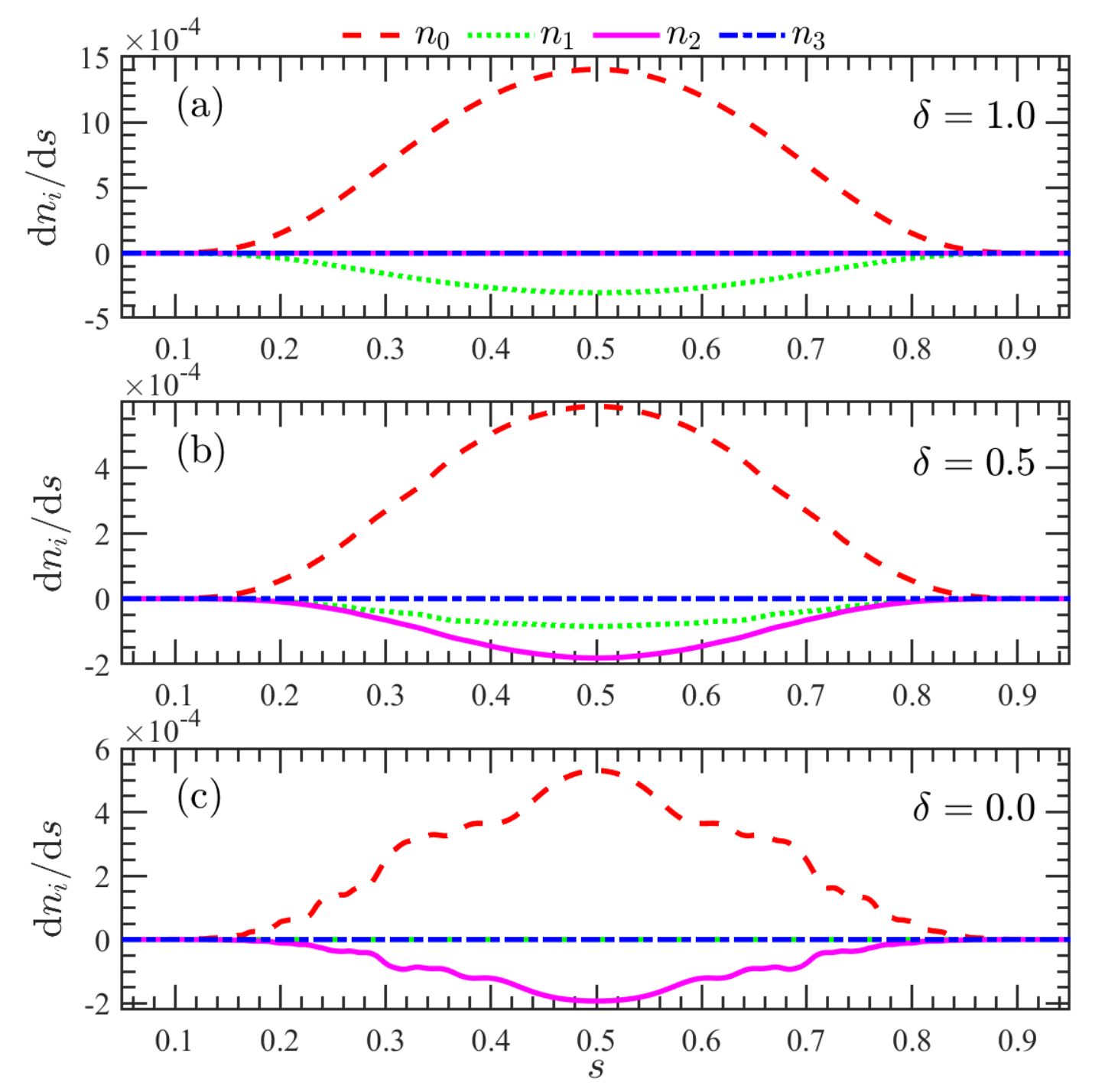}
\caption{The energy spectra of the produced positron via the NBW process in the head-on collision between a polarized photon and the laser pulse with different ellipticity: (a) $\delta=1$ circular polarization, (b) $\delta=0.5$ elliptical polarization, and (c) $\delta=0$ linear polarization.
The contributions from an unpolarized photon $n_{0}$ and those $n_{i}$ coupling to the photon polarization $\Gamma_{i}$ are compared.
The energy of the seed photon is $16.5~\trm{GeV}$.
The laser pulse has the intensity $\xi=1$, central frequency $\omega=4.65~\trm{eV}$ and the deflection angle $\theta=0$.}
\label{energy spectrums}
\end{figure}

In Fig.~\ref{energy spectrums}, we present the energy spectra of the produced positrons in the laser backgrounds with the same intensity $\xi=1$ but different ellipticity $\delta=1,~0.5,~0$ in Figs.~\ref{energy spectrums}~(a),~(b) and~(c) respectively.
As shown, the potential contributions coupling to the photon polarization are indeed appreciable for the total positron yield.
For the circularly polarized laser background, $\delta=1$ in (a) with $(\varsigma_1,~\varsigma_2,~\varsigma_3)=(0,~0,~1)$,
the relative importance of the contribution $n_{3}$, coupling to the circular polarization $\Gamma_{3}$ of the seed photon, is about $n_{3}/n_{0} \approx 22.3\%$ compared to the unpolarized contribution $n_{0}$.
The contributions $n_{1,2}$ coupling to the photon's linear polarization are zero, because the background field has no linear polarization~\cite{Tang2022PRD}.
By increasing the linear polarization of the background field $(\varsigma_1,~\varsigma_2,~\varsigma_3)=(0.6,~0,~0.8)$ in (b) with the ellipticity $\delta=0.5$,
the polarized contribution $n_{1}$ becomes important with $n_{1}/n_{0} \approx 27.8\%$,
while the importance of the polarized contribution $n_{3}$ decreases to about $n_{3}/n_{0} \approx 14.5\%$.
For the laser pulse with the full linear polarization in (c) with $\delta=0$ and $(\varsigma_1,~\varsigma_2,~\varsigma_3)=(1,~0,~0)$,
the polarized contribution $n_{3}$ becomes zero, and the relative importance of the polarized contribution $n_{1}$ increases to about $n_{1}/n_{0} \approx 32.6\%$.
With the decrease of the laser ellipticity, the harmonic structure becomes more clear in the energy spectra and appears around $s_{n>5} = \{1\pm [1-(2+\xi^2)/(n\eta)]^{1/2}\}/2$ when $\delta=0$~\cite{Tang2022PRD}.

\begin{figure}[t!!!]
   \includegraphics[width=0.45\textwidth]{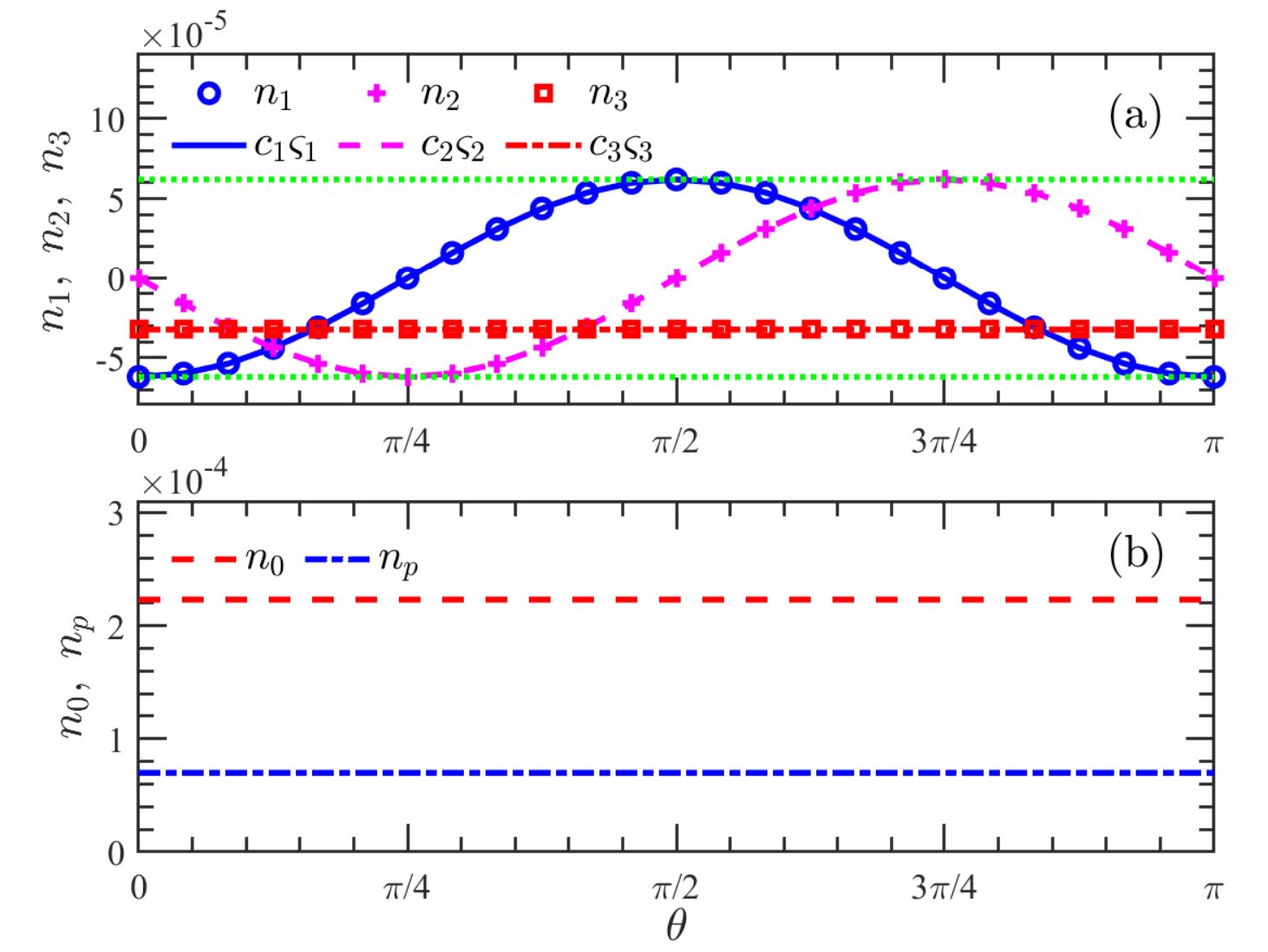}
   \caption{Different contributions to the positron yield of the NBW process in the elliptically polarized laser pulse with $\delta=0.5$ and the deflection angle $\theta\in[0,\pi]$.
   (a) The variation of the polarization contributions $n_{1,2,3}$ with the change of the field deflection angle. The full QED results (`cycle', `plus' and `square') are fitted with the corresponding laser Stokes parameters as $c_{1,2,3}\varsigma_{1,2,3}$, where $c_{1}=n_{1}(\theta=0)/\varsigma_{1}(\theta=0)$, $c_{2}=n_{2}(\theta=\pi/4)/\varsigma_{2}(\theta=\pi/4)$, and $c_{3}=n_{3}(\theta=0)/\varsigma_{3}(\theta=0)$. The green dotted lines denote the amplitude of the linear polarization contribution, \emph{i.e.}~$\pm (n^2_{1} + n^2_{2})^{1/2}$.
   (b) The unpolarised contribution $n_{0}$ and the maximized polarization contribution $n_{p}$ in~(\ref{Eq_M_Yiled}) from the seed photon with the optimal polarization in~(\ref{Eq_opti_polar}).
   The other parameters are the same as in Fig.~\ref{energy spectrums}.}
  \label{Fig_var_theta}
\end{figure}

In Fig.~\ref{energy spectrums}, the contribution $n_{2}$ is always zero with the change of the laser ellipticity $\delta$.
This is because the laser has no polarization preponderance along the direction of $\theta=\pi/4$, \emph{i.e.} $\varsigma_{2}=0$. 
To see the effect of the field deflection angle~$\theta$, we plot the variation of the polarization contributions $n_{i}$ with the change of $\theta$ in \mbox{Fig.~\ref{Fig_var_theta} (a)} for $\xi=1$ and $\delta=0.5$.
As shown, the polarization contributions $n_{1,2}$ vary in the trend as $(n_{1},~n_{2})\propto -(\cos2\theta,\sin2\theta)$ and $n_{3}$ is unchanged for different $\theta$.
All are in the same trend as the variation of the corresponding laser Stokes parameters $\varsigma_{1,2,3}$ in~(\ref{Eq_laser_Stokes}).
However, we also note that the amplitude of the linearly polarized contribution $(n^2_{1} + n^2_{2})^{1/2}$ is constant with the change of $\theta$ shown as the green dotted lines in \mbox{Fig.~\ref{Fig_var_theta} (a)}.
Therefore, the maximized polarization contribution $n_{p}$ in~(\ref{Eq_M_Yiled}) from the optimized polarization~(\ref{Eq_opti_polar}) is independent on the field's deflection angle $\theta$ as shown in \mbox{Fig.~\ref{Fig_var_theta} (b)},
in which we also find that the unpolarized contribution $n_{0}$ is unchanged for different $\theta$.
This is because of the azimuthal symmetry of the interaction geometry.
We can thus conclude that, for laser pulses with the fixed ellipticity $\delta$ and intensity $\xi$,
the field's deflection angle $\theta$ can only alter the relative value of the linear polarization contributions $n_{1,~2}$ with the constant amplitude $(n^2_{1} + n^2_{2})^{1/2}$, but not change the circularly polarized ($n_{3}$) and unpolarized ($n_{0}$) contributions.
To show the correlation between the polarization contribution $n_{i}$ and the corresponding laser Stokes parameter $\varsigma_{i}$,
we fit the numerical results in Fig.~\ref{Fig_var_theta} (a) respectively as \mbox{$n_{1}:~n_{1}(\theta=0)/\varsigma_{1}(\theta=0)\varsigma_{1} $}, $n_{2}:~ n_{2}(\theta=\pi/4)/\varsigma_{2}(\theta=\pi/4)\varsigma_{2}$, and $n_{3}:~n_{3}(\theta=0)/\varsigma_{3}(\theta=0)\varsigma_{3}$, and find the precise agreement between the numerical results and data fitting.

\begin{figure}[t!!!]
\includegraphics[width=0.45\textwidth]{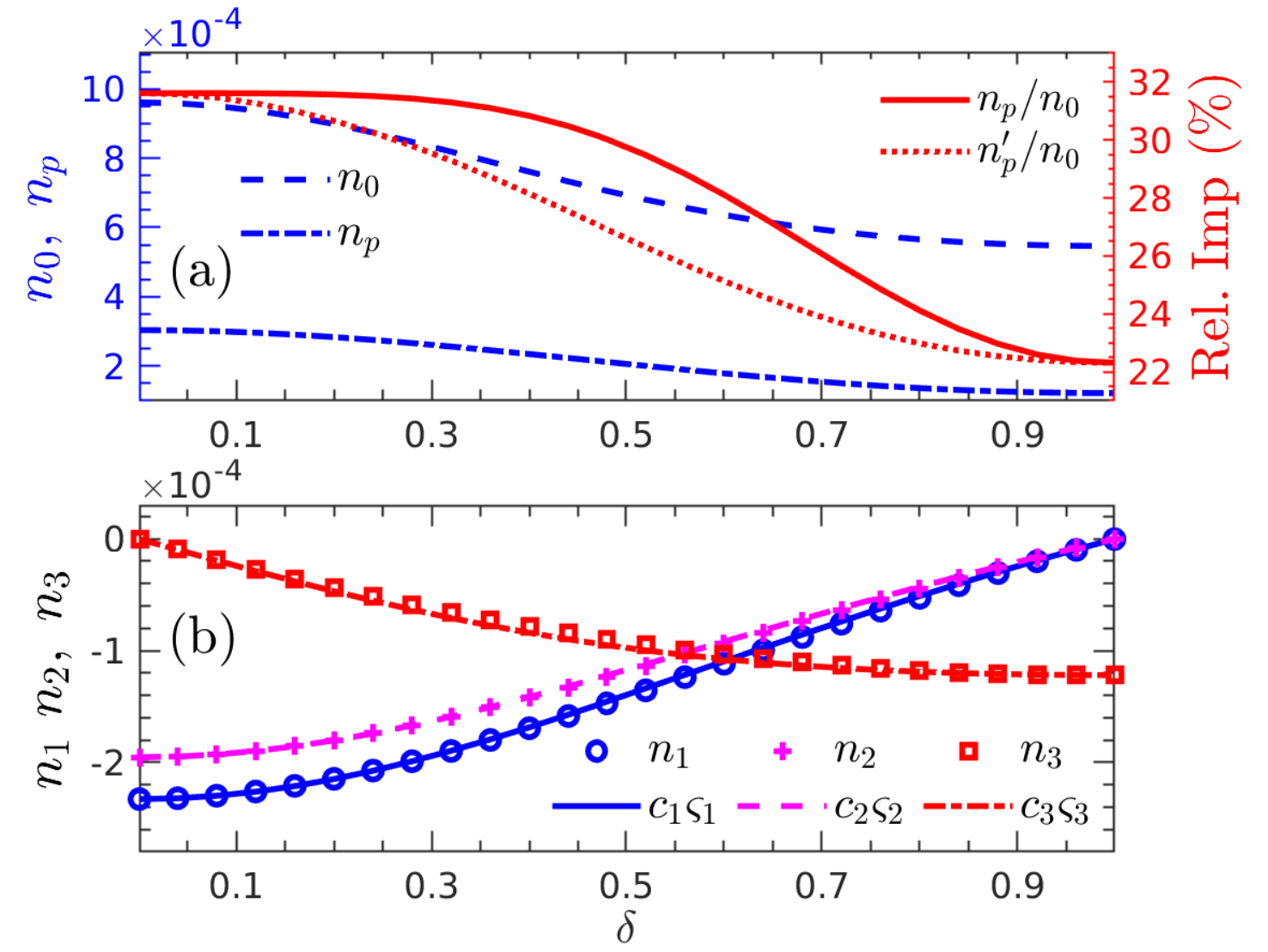}
\caption{Different contributions to the positron yield of the NBW process in the laser pulse with different ellipticity $\delta\in[0,~1]$, but the fixed laser power density $I= \xi^{2}(1+\delta^2)/2=1$ and deflection angle $\theta=\pi/9$.
(a) The unpolarized contribution $n_{0}$ and the maximized polarization contribution $n_{p}$ from the seed photon with the optimal polarization in~(\ref{Eq_opti_polar}).
The relative importance $n_p/n_0$ of the maximal polarization contribution $n_p$ is also plotted and compared with that of the polarization contribution $n'_{p}=-(\varsigma_{1}n_1+\varsigma_{2}n_2+\varsigma_{3}n_{3})$ from the photon state orthogonal to the laser polarization.
   (b) The variation of the polarization contributions $n_{1,2,3}$ with the change of the laser ellipticity.
   The full QED results (`cycle', `plus' and `square') are fitted with the corresponding laser Stokes parameters as $c_{1,2,3}\varsigma_{1,2,3}$, where $c_{1,2}=n_{1,2}(\delta=0)/\varsigma_{1,2}(\delta=0)$ and $c_{3}=n_{3}(\delta=1)/\varsigma_{3}(\delta=1)$.
   The laser power density $I=1$ corresponds to the real power density $I\approx 3.84\times 10^{19}~\trm{Wcm}^{-2}$.
   The other parameters are the same as in Fig.~\ref{energy spectrums}.}
   \label{Fig4_delta_I1}
\end{figure}

In Fig.~\ref{Fig4_delta_I1}, we show the variation of the different contributions to the positron yield with the change of the laser ellipticity $\delta$ for the fixed deflection angle $\theta=\pi/9$ and laser power density $I=1$, corresponding to $3.84\times 10^{19}~\trm{W}/\trm{cm}^{2}$.
As shown in Fig.~\ref{Fig4_delta_I1} (a), both the unpolarized contribution $n_{0}$ and the maximized polarization contribution $n_p$ from the optimal polarization~(\ref{Eq_opti_polar})~[shown in Fig.~\ref{Fig_Delta_dependence1} (b)] decrease with the increase of the laser ellipticity $\delta$ from $0$ to $1$.
This is because of the decrease of the field intensity $\xi=[2I/(1+\delta^2)]^{1/2}$.
Simultaneously, the relative importance, $n_{p}/n_{0}$, of the maximized polarization contribution decreases from about $31.6\%$ at $\delta=0$ for a linearly polarized laser pulse to about $22.3\%$ at $\delta=1$ for the laser pulse with pure circular polarization.
For comparison, we also plot the importance of the polarization contribution $n'_{p}=-(\varsigma_{1}n_1+\varsigma_{2}n_2+\varsigma_{3}n_{3})$ from the orthogonal state of the laser polarization,
which is clearly smaller than that from the optimal polarization state especially for the elliptically polarized laser with $\delta\approx0.5$.
In \mbox{Fig.~\ref{Fig4_delta_I1} (b)},
we see that the amplitude of the linear polarization contributions $n_{1,2}$ decrease with the increase of $\delta$,
while the amplitude of the contribution from the circular polarization, $n_{3}$, increases.
These variation are again in the same trend as the laser Stokes parameters in~(\ref{Eq_laser_Stokes}).
The difference between the two linear polarization contributions can be depicted as \mbox{$n_{1}/n_{2} \approx \tan2\theta=\varsigma_{1}/\varsigma_{2}$}.
The numerical results in \mbox{Fig.~\ref{Fig4_delta_I1} (b)} are respectively fitted as \mbox{$n_{1,2}:~n_{1,2}(\delta=0)/\varsigma_{1,2}(\delta=0)\varsigma_{1,2}$} and \mbox{$n_{3}:~n_{3}(\delta=1)/\varsigma_{3}(\delta=1)\varsigma_{3}$},
and again, we see the agreement between the numerical results and data fitting.
The slight difference around $\delta\approx 0.4$ implies the dependence of the polarization coupling between the seed photon and laser pulse on the laser ellipticity, as we will see later.

In this section, we investigate the NBW process in the laser pulse with the ellipticity $\delta\in [0,1]$ and deflection angle $\theta\in[0,\pi]$.
For the laser pulse with the ellipticity $\delta\in [-1,0]$,
the laser field would rotates in the opposite direction as the laser with the ellipticity $-\delta$ (see the expression for $\varsigma_{3}$).
The calculations would be consistent with the above results, except that the polarized contribution $n_{3}$ would change sign, but keeps the same amplitude.
For the laser pulse with the deflection angle $\theta\in[-\pi,0]$,
all the above results would also be the same except the polarized contribution $n_{2}$ would change sign because of the odd property of $\varsigma_{2}$.
All the calculations have be done for a relative long laser pulse, and for a ultra-short laser pulse, the conclusion would be different.

\subsection{Polarization coupling coefficients}

To manifest the dependence of the polarization contributions on the laser Stokes parameters, we consider an elliptically polarized monochromatic field with $f(\phi)=1$ in~(\ref{Vector potential}).
This is a good approximation for the pulses with slowly-varying envelope $f'(\phi)\approx0$.
After integrating the transverse momenta in~(\ref{NBW probability}), we can acquire the polarization contributions as
\begin{align}
(n_{1},n_{2},n_{3}) = \alpha I (\kappa_{nl}~\varsigma_{1},\kappa_{nl}~\varsigma_{2},\kappa_{nc}~\varsigma_{3})
\label{Eq_polar_cont_NBW}
\end{align}
where
\begin{subequations}
\begin{align}
\kappa_{nl}&= \frac{1}{\pi \eta}\hat{T}\sin\left(\frac{\vtheta\Lambda}{2\eta t s}\right)~g(\vtheta,\vphi)\,,\\
\kappa_{nc}&= \frac{1}{\pi \eta}\hat{T}\cos\left(\frac{\vtheta\Lambda}{2\eta t s}\right) h_{s}\left(\trm{sinc}^2\frac{\vtheta}{2} - \trm{sinc}\vtheta\right)\vtheta
\end{align}
\label{Eq_coupling_coeff_NBW}
\end{subequations}
are the coupling coefficients between the polarization of the seed photon and that of the laser pulse in the NBW process, and
\begin{align}
g(\vtheta,\vphi)= &\cos\vtheta +  \trm{sinc}^{2}\frac{\vtheta}{2}  - 2 \trm{sinc}~\vtheta \nonumber\\
                + &\frac{1}{\varsigma_{l}}\left(1 + \trm{sinc}^{2}\frac{\vtheta}{2} - 2\trm{sinc}~\vtheta \right)\cos2\vphi\nonumber\,.
\end{align}
$\Lambda$ is the Kibble mass and expressed as~\cite{kibble64}
\[\Lambda =1 + I - I \trm{sinc}^{2}\frac{\vtheta}{2}- I\varsigma_{l} \cos 2\vphi\left(\trm{sinc}^{2}\frac{\vtheta}{2} - \trm{sinc}\vtheta  \right)\]
depending on the laser power density $I=\xi^2(1+\delta^2)/2$ and its linear polarization degree $\varsigma_{l}$.
$\hat{T}$ is the integral operator given as
\[\hat{T}= \int^{1}_{0} \ud s \int^{\infty}_{-\infty} \ud\vphi\int^{\infty}_{0} \frac{\ud\vtheta}{\vtheta}\,,\]
with the average phase $\vphi=(\phi_{1}+\phi_{2})/2$ and the interference phase $\vtheta=\phi_{1}-\phi_{2}$~\cite{dinu16,seipt2017volkov,king19a}.

As we can see, in the NBW process, the polarization contribution $n_{i}$ is also proportional directly to the corresponding laser Stokes parameter $\varsigma_{i}$, as shown in Figs.~\ref{Fig_var_theta} and~\ref{Fig4_delta_I1}, with the coupling coefficients in~(\ref{Eq_coupling_coeff_NBW}) depending not only on the laser power, but also on the field ellipticity.
The two linear polarization components share, again, the same coupling coefficient because of the symmetry of rotating the linear polarization axis as discussed in Fig.~\ref{Fig_var_theta}.
We put the fine structure constant $\alpha$ out of the coupling coefficients as the NBW process is a single-vertex process,
and $I$ is because of the increase of the contributions with the laser power and in the perturbative regime, $n_{i} \propto \xi^2$ in~(\ref{Perturbative probability}).

\begin{figure}[t!!!]
\includegraphics[width=0.45\textwidth]{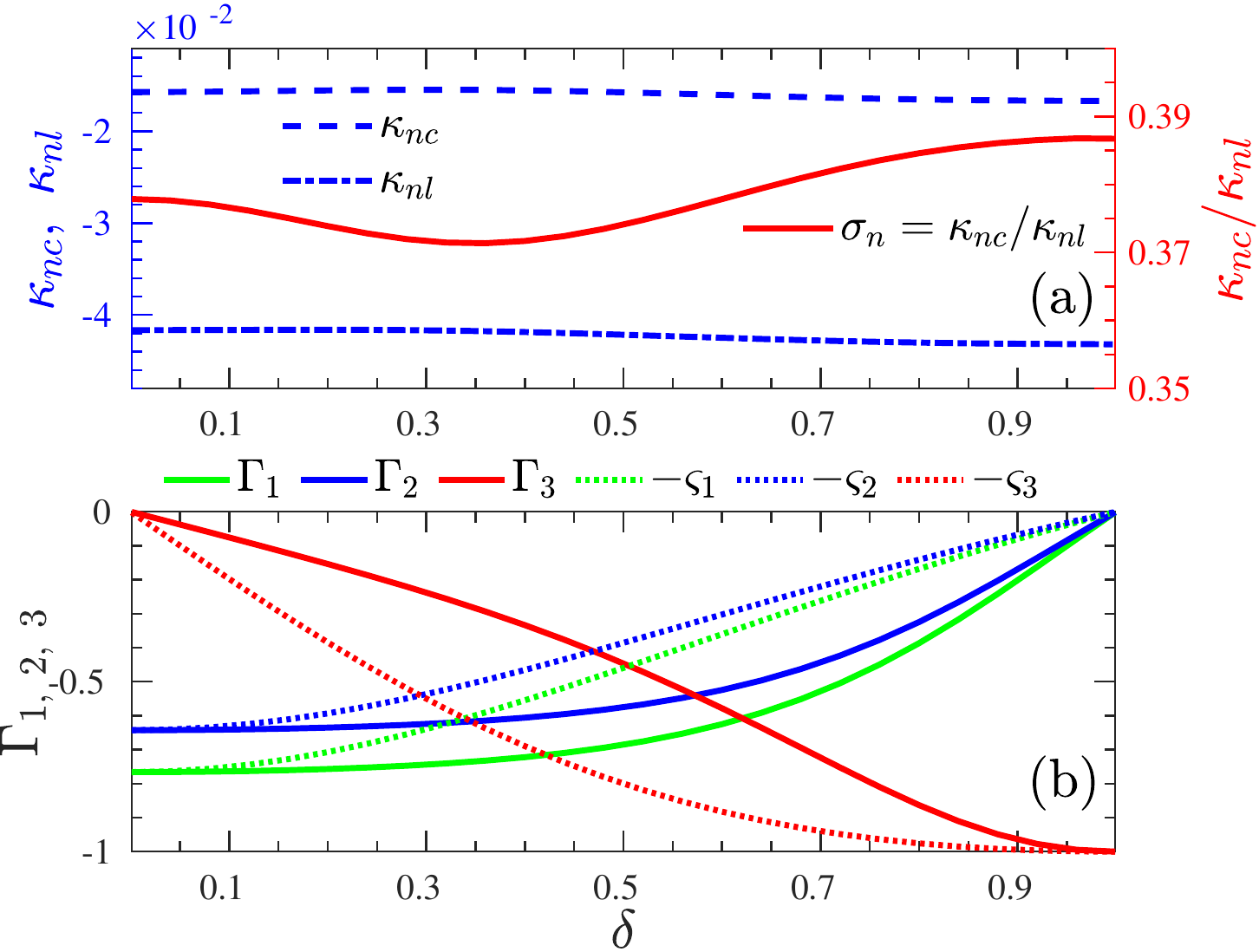}
\caption{(a) The variation of the coupling coefficients $\kappa_{nl},~\kappa_{nc}$ with the change of the field ellipticity. The ratio $\sigma_{n}=\kappa_{nc}/\kappa_{nl}$ is also plotted with the right $y$-axis.
The coefficient $\kappa_{nl}$ calculated from $n_{1}$ is exactly the same as that from $n_{2}$.
(b) The Stokes parameters of the photon's optimal polarization in (\ref{Eq_M_Yiled}) for different $\delta$. We also show the comparison with the orthogonal state, $-(\varsigma_{1},\varsigma_{2},\varsigma_{3})$ of the laser polarization.
The same parameters in Fig.~\ref{Fig4_delta_I1} are used.}
\label{Fig_Delta_dependence1}
\end{figure}
In Fig.~\ref{Fig_Delta_dependence1} (a), we present the dependence of the coupling coefficients $\kappa_{nl}$ and $\kappa_{nc}$ on the field ellipticity for the lasers with the fixed power $I=1$ and relatively long duration.
As shown, the value of $\kappa_{nl}$ and $\kappa_{nc}$ vary slightly with the change of the field ellipticity $\delta$,
and there exists significant difference between $\kappa_{nl}$ and $\kappa_{nc}$ with the ratio $\kappa_{nc}/\kappa_{nl}<1$, which also changes for different $\delta$.

\begin{figure}[t!!!]
\includegraphics[width=0.45\textwidth]{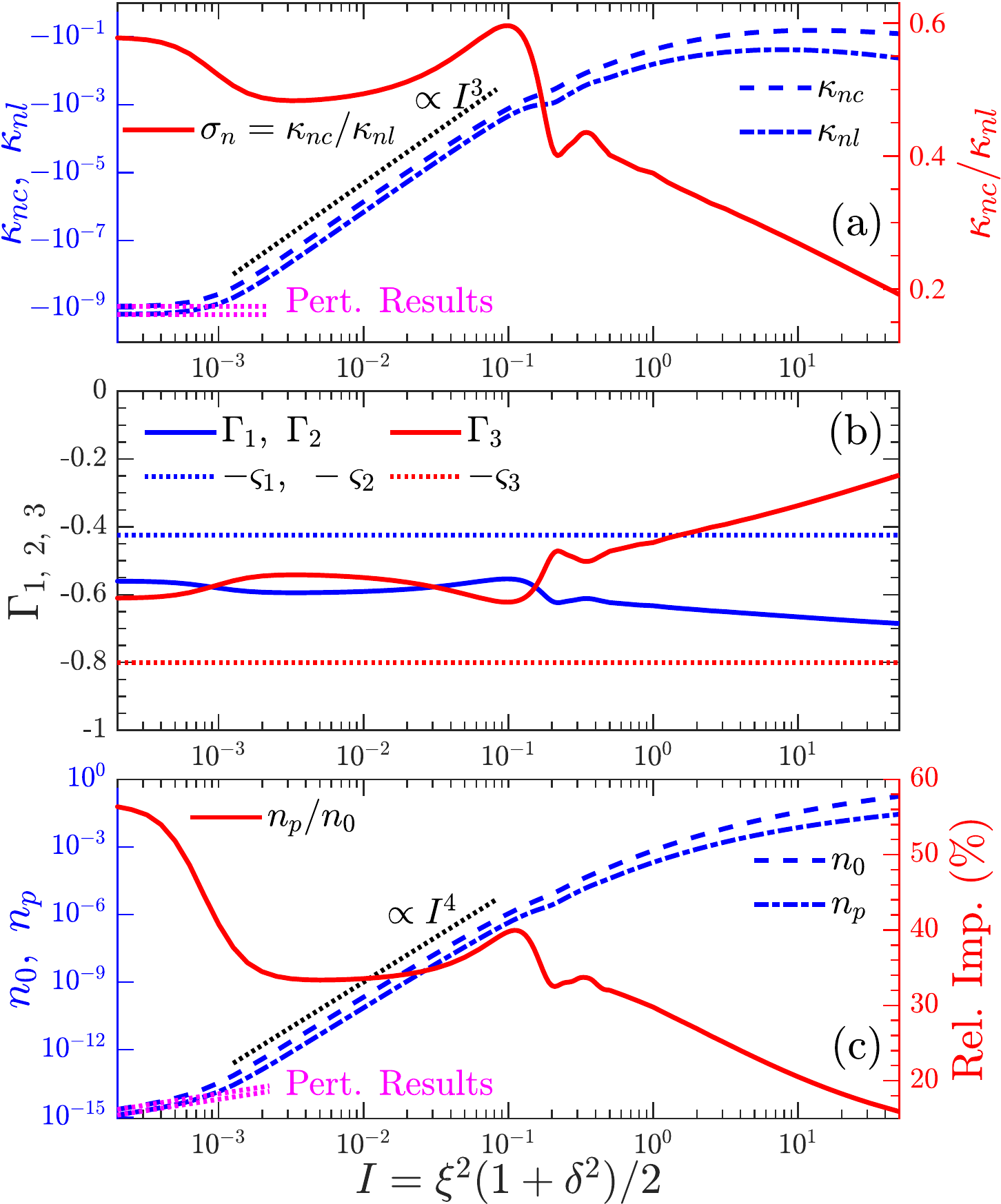}
\caption{(a) The variation of the coupling coefficients $\kappa_{nl},~\kappa_{nc}$ with the increase of the laser power density.
The dependence of the ratio $\sigma_{n}=\kappa_{nc}/\kappa_{nl}$ on the laser power is also presented with the right $y$-axis.
(b) The Stokes parameters of the photon's optimal polarization with the change of the laser power. $\Gamma_{1}=\Gamma_{2}$ as the field deflection angle is $\theta=\pi/8$.
(c) The yield from the unpolarized contribution $n_0$ and the maximal polarization contribution $n_p$, and the relative importance of the polarization effect $n_p/n_0$. 
In (a) and (b), the pink dotted lines are the corresponding perturbative results acquired from~(\ref{Perturbative probability}), and the black dotted lines show the varying trend of the curves.
The field ellipticity is $\delta=0.5$.
The other parameters are the same as in Fig.~\ref{Fig4_delta_I1}.}
\label{Fig_I_dependence}
\end{figure}

The dependence of $\kappa_{nl}$ and $\kappa_{nc}$ on the laser power density is presented in~\mbox{Fig.~\ref{Fig_I_dependence} (a)} for the fixed field ellipticity $\delta=0.5$ and deflection angle $\theta=\pi/8$.
As shown, 
in the low-power density region $I<10^{-3}$, $\kappa_{nl}$ and $\kappa_{nc}$ are independent on the laser power $I$ because the LBW process dominates the production, $\kappa_{nl}$ and $\kappa_{nc}$ can be acquired alternatively from the perturbative result~(\ref{Perturbative probability}) with $\kappa_{l}$ and $\kappa_{c}$ depending only on the parameter $\beta$.
The value of $\kappa_{nl}$ and $\kappa_{nc}$ are determined by the energy parameter $\eta$ and the pulse envelope.
In this region, 
the positron yield increases as $n_0, n_p\propto I$ shown in Fig.~\ref{Fig_I_dependence} (c) because of the single-photon effect with the high-frequency components from the finite-pulse effect~\cite{TangPRD096019}.
In the intermediate laser power region,~\mbox{$10^{-3}<I<10^{-1}$}, the coupling coefficients increase as $\kappa_{nl}, \kappa_{nc} \propto I^{3}$ because of the multiphoton perturbative effect, in which $4=\lceil 2/\eta \rceil$ laser photons are involved in the production process and
the positron yield increase in the trend as $n_{0}, n_p\propto I^{4}$ in \mbox{Fig.~\ref{Fig_I_dependence} (c)},
where $\lceil x \rceil$ denotes the minimal integer larger than $x$.
With the further increase of the laser power, $I\gtrsim 0.5$, this $4$-photons channel is forbidden and a higher number of laser photons, $n=\lceil 2(1+I)/\eta \rceil$, would be involved in the production process.
Therefore, the fully non-perturbative effect would be dominant.
The increase of the coupling coefficients $\kappa_{nl}$~and~$\kappa_{nc}$ become slower, as well as the increase of the positron yield in~\mbox{Fig.~\ref{Fig_I_dependence} (c)}.
In \mbox{Fig.~\ref{Fig_I_dependence} (a)}, we can also see the evident difference between $\kappa_{nl}$ and $\kappa_{nc}$ in the broad laser power region with the ratio $\kappa_{nc}/\kappa_{nl}<1$ depending also sensitively on the laser power.
This difference would result in the deviation of the optimal photon polarization from the completely orthogonal state of the laser polarization.

\subsection{Optimal photon polarization}
From~(\ref{Eq_coupling_coeff_NBW}), the optimal polarization of the seed photon~(\ref{Eq_opti_polar}) can be written as
\begin{align}
(\Gamma_{1},~\Gamma_{2},~\Gamma_{3})=\hat{\kappa}_{nl}\frac{(\varsigma_{1},~\varsigma_{2},~\sigma_{n} \varsigma_{3})}{(\varsigma_{1}^2 + \varsigma_{2}^2 + \sigma^2_{n}~\varsigma_{3}^2)^{1/2}}\,,
\label{Eq_opti_polar_NBW}
\end{align}
based on the polarization of the laser pulse, where \mbox{$\hat{\kappa}_{nl}=-1$} is the sign of $\kappa_{nl}$ acquired numerically,
and $\sigma_{n}=\kappa_{nc}/\kappa_{nl}$ denotes the difference between the coupling coefficients $\kappa_{nl}$ and $\kappa_{nc}$.
If $\sigma_{n}\neq 1$, the photon's optimal polarization state would deviate from the orthogonal state $-(\varsigma_{1},\varsigma_{2},\varsigma_{3})$ of the laser polarization.

As shown in Fig.~\ref{Fig_Delta_dependence1} (a), $\sigma_{n}$ is much smaller than $1$ for different $\delta$. 
Therefore, the optimal polarization state of the seed photon, for the maximal yield, is much different from the orthogonal state $-(\varsigma_{1},\varsigma_{2},\varsigma_{3})$ of the laser polarization as one can see in Fig.~\ref{Fig_Delta_dependence1} (b),
except in the regions around $\delta\approx 0,~1$, where the laser is linearly and circularly polarized, respectively.
With the optimized photon polarization in Fig.~\ref{Fig_Delta_dependence1} (b), the production yield could be enhanced for more than $20\%$ compared to the unpolarized case as shown in Fig.~\ref{Fig4_delta_I1} (a).

In Fig.~\ref{Fig_I_dependence} (b), the optimal polarization state of the seed photon is presented in a broad laser power region for the specified ellipticity $\delta=0.5$. 
Because the field deflection angle is $\theta=\pi/8$, the two linear polarization components are equal,~$\Gamma_{1}=\Gamma_{2}$.
Again, because of the evident difference between $\kappa_{nl}$ and $\kappa_{nc}$ in Fig.~\ref{Fig_I_dependence} (a),
the photon's optimal polarization state deviates considerably from the orthogonal state of the laser polarization as shown in Fig.~\ref{Fig_I_dependence} (b).
Especially in the non-perturbative regime $I>0.5$, the circular polarization degree $|\Gamma_{3}|$ of the optimal polarization decreases rapidly with the increase of $I$,
because of the rapid decrease of the ratio $\kappa_{nc}/\kappa_{nl}$ for larger $I$ in Fig.~\ref{Fig_I_dependence} (a),
which means that the contribution from the circular polarization becomes less important. 
In the ultra-high intensity regime $\xi\gg10$ (not shown in Fig.~\ref{Fig_I_dependence}), in which the locally constant field approximation would work precisely~\cite{king19a,Tang2022PRD},
the contribution from the circular polarization would be negligible, \emph{i.e.}~$k_{nc}\to0$ and $\Gamma_{3}\to0$.
This is because the formation length of the NBW process becomes much shorter than the typical length of the field variation~\cite{ritus85} and the laser pulse would work as a linearly polarized field with the  direction varying with the laser phase~\cite{Tang2022PRD}.

With the polarization-optimized seed photon, the positron yield could be enhanced appreciably as shown in Fig.~\ref{Fig_I_dependence} (c).
In the perturbative intensity region $I<10^{-3}$, the positron yield could be enhanced more than $55\%$ by the polarization effect compared with the unpolarized case,
and in the multi-photon perturbative region~\mbox{$10^{-3}<I<10^{-1}$}, the yield enhancement is about $34\%$ from the optimized polarization state.
With the further increase of the laser power, even though the relative importance of the polarization contribution becomes less,
the positron yield could still be improved for more than $16\%$ at $I \lesssim 50$.

\section{Conclusion}~\label{sec_conc}
The optimization of the photon polarization state to the maximal positron yield of the Breit-Wheeler pair production is investigated in arbitrarily polarized plane wave backgrounds for a broad intensity region.
Both the polarization of the photon and the laser pulse are comprehensively described with the classical Stokes parameters.

The optimal polarization state of the seed photon is resulting from the polarization coupling with the laser pulse/photon in the production process.
For the laser pulse with the pure linear or circular polarization, the seed photon's optimal polarization is the orthogonal state of the laser pulse. However, because of the evident difference between the coupling coefficients for the linear and circular polarization components, 
the seed photon's optimal polarization state in elliptically polarized laser backgrounds, deviates considerably from the orthogonal state of the laser polarization, especially in the ultrahigh-intensity regime in which the linear-polarization coupling coefficient is much larger than that of the circular polarization and thus the seed photon's optimal polarization would tend to the linear polarization.

With the polarization-optimized seed photon, the positron yield could be considerably enhanced in a broad intensity region. For the laser intensity region,~\mbox{$\xi\sim\mathcal{O}(1)$}, of current laser-particle experiments, the yield enhancement from the optimized photon polarization could be more than $20\%$ compared to the unpolarized case.

\section{Acknowledgments}
The author thank A. Ilderton for helpful suggestions and comments on the manuscript.
The author acknowledge the support from the National Natural Science Foundation of China, Grant No.12104428.
The work was carried out at Marine Big Data Center of Institute for Advanced Ocean Study of Ocean University of China.

\bibliographystyle{apsrev}
\bibliography{NBW_optiaml_polar}

\end{document}